\documentclass[prl,twocolumn,amsmath,amssymb,longbibliography,showpacs,superscriptaddress]{revtex4-1}
\usepackage{graphicx,color}
\usepackage{latexsym}
\usepackage{pgfplots}
\usepackage{empheq}

\newcommand{\1}{{\scriptscriptstyle{1}}}

\renewcommand{\vec}[1]{\boldsymbol{#1}}

\newcommand{\caD}{{\mathcal D}}

\newcommand{\caH}{{\mathcal H}}

\newcommand{\caO}{{\mathcal O}}

\newcommand{\caT}{{\mathcal T}}

\newcommand{\diff}{d} 

\newcommand{\mean}[1]{{\left< #1 \right>}}
\newcommand{\genL}{{\mathbb L}}

\begin{document}

\title{Temperature response in nonequilibrium stochastic systems}

\author{Gianmaria Falasco}
\email{falasco@itp.uni-leipzig.de}
\affiliation{Institut f\"ur Theoretische Physik, Universit\"at Leipzig,  Postfach 100 920, D-04009 Leipzig, Germany}

\author{Marco Baiesi}
\email{baiesi@pd.infn.it}
\affiliation{Department of Physics and Astronomy, University of Padova, Via Marzolo 8, I-35131 Padova, Italy}
\affiliation{INFN, Sezione di Padova, Via Marzolo 8, I-35131 Padova, Italy}

\date{\today}

\begin{abstract}
The linear response to temperature changes is derived for systems with overdamped stochastic dynamics.
Holding both in transient and steady state conditions, the results allow to compute nonequilibrium thermal susceptibilities from unperturbed correlation functions. These correlations contain a novel form of entropy flow due to temperature unbalances, next to the standard entropy flow of stochastic energetics and to complementary time-symmetric dynamical aspects. Our derivation hinges on a time rescaling, which is a key procedure for comparing apparently incommensurable path weights. An interesting notion of thermal time emerges from this approach.
\end{abstract}

\pacs{05.40.-a,
      05.70.Ln}

\maketitle

In thermodynamic equilibrium,  the linear response coefficients for perturbing forces (e.g conductivity as a response to
an electric potential) and perturbed temperatures (e.g.~thermal expansion coefficients or specific heats) may be
computed with the fluctuation-dissipation theorem. So-called Kubo formulas relate the response to the unperturbed correlation between the observable and the entropy produced by the perturbation~\cite{kub66}. 
Out of equilibrium such a clear picture is lacking yet.

For nonequilibrium systems the standard linear response to deterministic forcing has been developed via many approaches (see e.g.~\cite{fal90,cug94,rue98,nak08,che08,spe09,pro09,ver11a,lip05,lip07,bai09,bai09b}).
In comparison, there is a small number of results obtained for the response to temperature changes~\cite{che09b,bok11,bai14,bra15}. For example there is no formula to express, as a function of steady state unperturbed correlations, the thermal expansion coefficient for a solid kept in a temperature gradient (e.g.~as in experimental setups~\cite{agu15,con12} or in models of coupled oscillators~\cite{gre11,lep03}).
The construction of a general steady state thermodynamics~\cite{oon98,hat01,sek07,sag11,ber13,mae14,kom15,man15} needs at its heart the understanding of temperature responses, for example in defining nonequilibrium specific heats~\cite{bok11}. 
A nonequilibrium thermal response should as well be the subject of study in related fields, such as climatology~\cite{abr08,lac09,luc11}.

This paper introduces a  theory for the linear response to a change of the reservoirs' temperature. 
We consider nonequilibrium overdamped systems. Mesoscopic systems of this kind, including driven colloids~\cite{bli07a,kot10} and active matter~\cite{mar13}, 
are used as paradigms in the attempt to generalize equilibrium concepts, such as free energies, within the framework of a steady state thermodynamics~\cite{oon98,hat01,sek07,sag11,ber13,mae14,kom15,man15}. Moreover, 
they offer the possibility to observe experimentally the statistical fluctuations of energy fluxes~\cite{gom11,gom12}.

Our approach is inspired by a scheme based on path weighs and developed for the response to forces~\cite{bai09,bai09b}. For that theory the response turns out to be the sum of two terms, as in previous studies~\cite{cug94,lip05,lip07}). The first one is half of the unperturbed correlation $\mean{\caO S}$ between observable $\caO$ and entropy $S$ produced by the perturbation, i.e.~half of a Kubo formula. The second is the correlation $-\mean{\caO K/2}$, where the term $-K/2$, of which we still have a less intuitive grasp, is the time-symmetric part of the action weighting the system's trajectories. In order to avoid singularities emerging in a related temperature response~\cite{bai14} based on a time-discretization, we introduce a time rescaling that significantly changes the derivation. As a result, the susceptibility is written as sum of unperturbed correlations containing well-defined (stochastic) integrals. Moreover, an intriguing and unexpected picture emerges: in $S$, the heat fluxes as described in the context of stochastic energetics~\cite{oon98,sek10,har05} appear accompanied by a second form of entropy production (not present in~\cite{bai14}), which is relevant when the system is coupled to reservoirs at different temperatures.

The overdamped diffusive system we consider is described by $i=1,\ldots,N$ 
degrees of freedom $\vec x = \{x_i\}$, evolving in the unperturbed state as
\begin{align}\label{dotx}
\dot x_i(t) = F_i(\vec x(t)) + \sqrt{T_i}\, \xi_i(t),
\end{align}
where every Gaussian white noise $\xi_i$ is uncorrelated with the others,
$\mean{\xi_i(t)\xi_j(t')} = 2 \delta_{ij}\delta(t-t')$. A constant temperature $T_i$ is associated to each reservoir. 
For simplicity, in particular, we choose a subset $\caT$ so that $T_i=T$ if $i\in \caT$, which is then considered as a single heat bath to be perturbed. 
An indicator function, $\epsilon_i=1$ only if $i\in\caT$ and $\epsilon_i=0$ otherwise, is useful to keep track of the perturbed degrees of freedom~\footnote{Note that more general conditions on $\epsilon$ and $T$ may be imposed with the same formalism.}.
We seek the linear response of a generic state observable $\caO(\vec x)$ to a change in $T$, namely,
\begin{align}
R_{\caO T}(t,t')\equiv \frac 1 T \frac{\delta \mean{\caO(t)}^h}{ \delta h(t')} \bigg|_{h=0}
\end{align}
where $h(t)\ll 1$ is the modulation of the reservoir temperatures, $\Theta_i(t) = T_i[1+\epsilon_i h(t)]$.
Note that temperatures do not depend on the coordinates, hence there is no ambiguity in the interpretation of the stochastic equation. Throughout this paper we will always consider the Stratonovich convention, implying standard rules of functional calculus. The system may be brought far from equilibrium (a) by generic nonconservative forces $F_i$,
(b) by different $T_i$'s, and (c) by a relaxation from an initial transient condition. Indeed, given that the perturbation is turned on at times $t>0$, the initial density of states $\rho_0(\vec x)$ at $t=0$ may coincide or not with a steady state density. For economy of notation we do not recall this explicitly in the statistical averages, denoted by $\mean{\ldots}^h$ and $\mean{\ldots}$ for the perturbed ($h\ne 0$) and unperturbed ($h=0$) case, respectively.

 The backward generator associated to the Markovian dynamics \eqref{dotx} is
\begin{align}\label{L}
\genL = \sum_{j=1}^N \genL_j
\quad \textrm{with} \quad
\genL_j =  F_j(\vec x) \partial_{x_j} + T_j \partial^2_{x_j},
\end{align}
written in a notation that will be useful later. It will turn out that also the following
modified operator is useful to describe the temperature response, when the temperature $T$ of a reservoir is altered:
\begin{align}\label{L_T}
\genL^{(T)} \equiv \sum_{j=1}^N \frac T {T_j} \genL_j = \sum_{j=1}^N \left( \frac T {T_j}  F_j(\vec x) \partial_{x_j} + T \partial^2_{x_j}\right),
\end{align}
which acts on the observables as if all temperatures were equal to $T$ and all forces $F_j$ were rescaled by 
$T/T_j$. While $\genL$ gives the derivative of a state observable $\caO$ with respect to the kinematic time $t$, i.e. $\mean{\genL \caO}=\partial_t \mean{\caO}$,  $\genL^{(T)}$ gives the variation of each degree of freedom with respect to its own \emph{thermal} time $\tau_j \equiv t T_j/T$, namely,
\begin{align}
\mean{\genL^{(T)} \caO}=\mean{ \sum_j \frac{\diff x_j}{\diff \tau_j} \partial_{x_j}\caO}\equiv \partial_{t}^{(T)}\mean{\caO},
\end{align} 
such that a generalized time derivative results defined.

Before spelling out the derivation of our results, some physical insights on the meaning of a temperature change can be gained by performing the time transformation $T \diff \tau= \Theta(t) \diff t$ in~\eqref{dotx}. Taking $N=1$ for simplicity, upon perturbation~\eqref{dotx} becomes to first order in~$h$
\begin{align}\label{tau}
\dot x (\tau)
 &\simeq [1- h(\tau)] F(x(\tau)) + \sqrt{T}\, \xi(\tau),
\end{align}
which clearly shows that, in such time coordinate, a temperature perturbation is equivalent to a force perturbation. The response to a small decrease in $F$ is given by a theory of nonequilibrium linear response \cite{bai09b} as
\begin{align}\label{ROF}
R_{\caO F}(\tau, \tau')=-\frac{1}{2T} \mean{\caO(\tau) [\dot x(\tau') F(\tau')  - \dot K(\tau')]},
\end{align}
where $\dot K(x(\tau')) =F^2(x(\tau')) +T \partial_x F(x(\tau'))$.
If the system were in equilibrium, so as that $F= -\partial_x \caH$ with $\caH$ the system's Hamiltonian, one would expect the correlation functions to be invariant under a time reparametrization. Therefore, from~\eqref{ROF} the response to a temperature change in equilibrium is obtained as 
\begin{align}\nonumber
R_{\caO T}(t-t')=\frac{1}{2T^2} \mean{\caO(t) [\dot \caH (t')  - \genL \caH (t')]},
\end{align}
which is recognized as a Kubo formula, since in equilibrium $ \mean{\caO(t) \genL \caH (t')} = - \langle \caO(t) \dot \caH (t') \rangle $ \cite{bai09}. Out of equilibrium instead $R(\tau, \tau')$ depends implicitly on $h$ through the time variables and no further simplification of~\eqref{ROF} appears possible.

Nevertheless, the diffusive character of the system suggests to replace the above time change with the space coordinate change $y_i(t)/ \sqrt{T_i}={x_i(t)} / \sqrt{\Theta_i(t)}$~\cite{rie96}, so that~\eqref{dotx} reads
\begin{align}\label{doty0}
\dot y_i(t) = & \sqrt{\frac{T_i}{\Theta_i}} F_i\left(\vec x(\vec y)
 \right) 
 - \frac 1 2 y_i(t)\frac{\dot \Theta_i}{\Theta_i} + \sqrt{T_i}\, \xi_i.
\end{align}
Perturbed averages are now calculated with the path weights for the process $\vec y$, i.e the statistical weight $P^h[\vec y]$ of the trajectory ${\{\vec y(s): 0\leqslant s \leqslant t\}}$. In particular, for all times $0<t' < t$ and any state observable $\caO$, the linear response is evaluated as \footnote{If $\caO$ is a functional of the trajectory, i.e. it is extensive in time like, e.g., heat flows, the response contains the additional term $\mean{\frac{\delta \caO}{\delta h(t')}\big|_{h=0}}$.}
\begin{align}\label{ROT}
R_{\caO T}(t,t')
 = \int \caD \vec y \caO(\vec y(t)) \frac{\delta P^h[\vec y]}{\delta h(t')} \bigg|_{h=0}.
\end{align}
Here, only terms of order ${\rm O}(h)$ are needed, hence we can directly linearize the Langevin equation~\eqref{doty0} obtaining 
\begin{align}\label{doty}
&\dot y_i \simeq 
 F_i\left(\vec{y}\right) +\frac h 2  f_i\left(\vec{y}\right) 
- \frac {\epsilon_i} 2 \dot h y_i  + \sqrt{T_i}\, \xi_i,
\end{align}
where we recognize a standard perturbing force, 
\begin{align}
f_i =  \sum_{j=1}^N \epsilon_j y_j \partial_{y_j}  F_i - \epsilon_i F_i ,
\end{align}
 and a second one, $-\dot h \epsilon_i y_i/2$, which is atypical in that it is modulated by $\dot h(t)$. Note that in expanding the force $F_i$ it is implicitly required that the values of $y_i$ are bounded.

The path weight $P^h[\vec y]$ is obtained with a standard procedure from the Gaussian path weight for $\vec \xi$, regarding $\vec \xi$ as a functional of $\vec y$ via~\eqref{doty} \cite{zinn02}:
\begin{align}\label{pathint}
P^h[\vec y] \propto &  \prod_{i=1}^N \exp\Bigg\{ 
\!\!- \! \frac{1}{4 T_i}\int_0^t \! \! \diff s \big[\dot y_i-  F_i
- \frac 1 2 \big( h f_i - \epsilon_i \dot h y_i\big) \big]^2  \nonumber \\
& - \frac{1}{2} \int_0^t \!\!\diff s \big[
\partial_{y_i} F_i + \frac 1 2  \big(h \partial_{y_i} f_i -  \epsilon_i\dot h\big) 
\big] \Bigg\}
\end{align}
(the dependence of all terms on the time $s$ is understood).
Plugging~\eqref{pathint} in~\eqref{ROT} we encounter the modulation $\dot h$, which wraps in a time derivative  the standard result valid for deterministic perturbations, namely
$\frac{\delta}{\delta h(t')} \int_0^t \diff s \dot h [y_i (\dot y_i-F_i)] = -  \partial_{t'}[y_i (\dot y_i-F_i)] $.
Assuming that the interaction forces are two body potentials, so that $\partial_{y_j} F_i=\partial_{y_i} F_j$,
by massaging the formulas and using  $y_i |_{h=0}=x_i$ we finally derive
\begin{align}
\label{dP}
&\frac{\delta P^h[\vec y]}{\delta h(t')} \bigg|_{h=0}\!\!\! = 
\sum_i \frac{\epsilon_i}{4} \bigg\{ -\frac{2 F_i \dot x_i}{T_i}
+\frac{F^2_i}{T_i} 
-x_i \sum_j \frac {\genL_j F_i} {T_j}
 \nonumber \\
& + x_i \sum_j \left( \frac 1 {T_j} - \frac 1 {T_i}\right) \dot x_j\partial_{x_j} F_i
 +\frac{\partial^2_{t'} x_i^2}{2 T_i}
\bigg\}(t')  P[\vec x].
\end{align}
Given our choice $\epsilon_i=1$ for $i\in \caT$,
the response function of $\caO(t)$ to the variation of $T$ is thus written as
\begin{subequations}
\label{resp}
\begin{align}
\label{resp_a}
&R_{\caO T}(t,t')=  \sum_{i\in \caT} \frac{1}{4 T^2}\bigg\{ 
- 2 \mean{\caO(t) F_i(t')\dot x_i(t')}\\
\label{resp_b}
&\qquad \quad + \mean{\caO(t) \sum_{ j \notin \caT}
 \left( \frac T {T_j} - 1 \right) [ x_i \dot x_j\partial_{x_j} F_i](t')}\\
\label{resp_c}
&\qquad \quad+\mean{\caO(t) F^2_i(t')} -\mean{\caO(t) x_i(t') \genL^{(T)} F_i(t')}\\
\label{resp_d}
&\qquad \quad+\frac{1}{2}\partial^2_{t'} \mean{\caO(t)  x_i^2(t')}
\bigg\}.
\end{align}
\end{subequations}
In~\eqref{resp_a} $J_i = -F_i\dot x_i$ is the heat flux from the $i$-th bath \cite{sek10}. In addition, in~\eqref{resp_b} there appears a novel kind of heat flux, 
\begin{align}
&J_i^{\rm int}(t') =  x_i \sum_j \left( \frac T {T_j} - 1\right) \dot x_j\partial_{x_j} F_i\\
&\quad =  \left(\partial_{t'}^{(T)} - \partial_{t'} \right) (x_i F_i) 
 = -\sum_j\left(\frac{\diff x_j}{\diff \tau'} - \frac{\diff x_j}{\diff t'} \right)\partial_{x_j} V_i \,, \nonumber
\end{align}
which vanishes when kinematic and thermal times coincide, i.e.~when the system is isothermal previous to the perturbation. The virial  $V_i \equiv - x_i F_i$ of the $i$-th degree of freedom (whose average value equals $T_i=T$ even in a nonequilibrium steady state~\cite{in_preparation}) is seen to act as a potential energy for $x_j$.
The meaning of $J_i^{\rm int}$ is thus understood as the difference between the heat absorption rate in the kinematic time and that in the thermal time, generated by forces $\partial_{x_j} V_i$ on $x_j$'s. 
Thus, the total entropy flux from the selected heat bath to the system, $\sum_{i\in \caT} J_i/T$,
is side by side with the entropy flux $\sum_{i\in \caT} J_i^{\rm int}/T$ due to the heat currents installed within the system by the presence of different coupled temperature reservoirs. These two terms are time-antisymmetric, that is, they change sign by going through the trajectory backward in time. The remaining terms, namely \eqref{resp_c} and  \eqref{resp_d} contain the correlation between the observable and time symmetric quantities.

During the last decade there was an increase of interest in time-symmetric fluctuating 
quantities (see e.g.~\cite{lec05,mer05,mae06,gar09,bai09,elm10}), as it is becoming
clearer that they must complement entropy fluxes for a deeper understanding of statistical mechanics. 
For example, the dynamical activity of a jump process (counting the number of jumps) is a key aspect 
for characterizing glassy dynamics~\cite{lec05,mer05,gar09,elm10}.
In linear response it was found that the time-symmetric sector of path weights is often related to the mean tendency of the system to change the perturbing potential, e.g.~for jump processes it becomes a shift in escape rates~\cite{bai09,bai09b}. The adjective ``frenetic'' was used to label this property~\cite{bai09,bai09b,bai10}.

In \eqref{resp_c} we find frenetic contributions that do depend explicitly on forces of the system while the last term \eqref{resp_d} does not. The presence of such term is necessary for having a possibly non-zero response also for free diffusion. For example, for a free particle starting from $x(0)=0$ the theory  yields a response of the mean square displacement $\mean{x^2(t)}$ to an increase of $T$ which is correctly 
$R_{x^2T}(t,t') = \frac 1{8 T^2}\partial^2_{t'}\mean{x^2(t) x^2(t')} = 2$ 
(or more in general twice the mobility, if we had put such constant in front of the forces~\footnote{
Our results can be easily generalized if a mobility $\mu_i$ (the inverse of a damping constant) 
is associated with each degree of freedom: one just needs the replacements $F_i\to \mu_i F_i$ and 
$T_i \to \mu_i T_i$ in the formulas, besides for the $1/T$ prefactor in the definition of the response.
}; 
the calculation considers the Gaussian statistics and uses Wick's theorem).

Upon integration of \eqref{resp} we get the susceptibility
\begin{align}
\label{chi}
\chi_{\caO  T}(t) \equiv 
& \int_0^t \diff t' R_{\caO T}(t,t')
=\frac1 {2 T} \left[ \mean{\caO(t) S  } -  \mean{\caO(t) K} \right]
\end{align}
where $S$, the entropy change of reservoir $\caT$, 
contains the Stratonovich integrals of \eqref{resp_a} and \eqref{resp_b},
while the ``frenesy'' $K$ contains the remaining integrals of \eqref{resp_c} and \eqref{resp_d}.
In $-K$ in particular there appears a pair of boundary terms
$\frac 1 {4T}\left.\partial_{t'}\mean{ \sum_{i\in \caT}x_i^2(t') \caO(t)}\right|_{t'=0}^{t'=t}$
in which left derivatives are performed to keep $t'\le t$.

\begin{figure}[!t] 
\includegraphics[width=0.98\columnwidth]{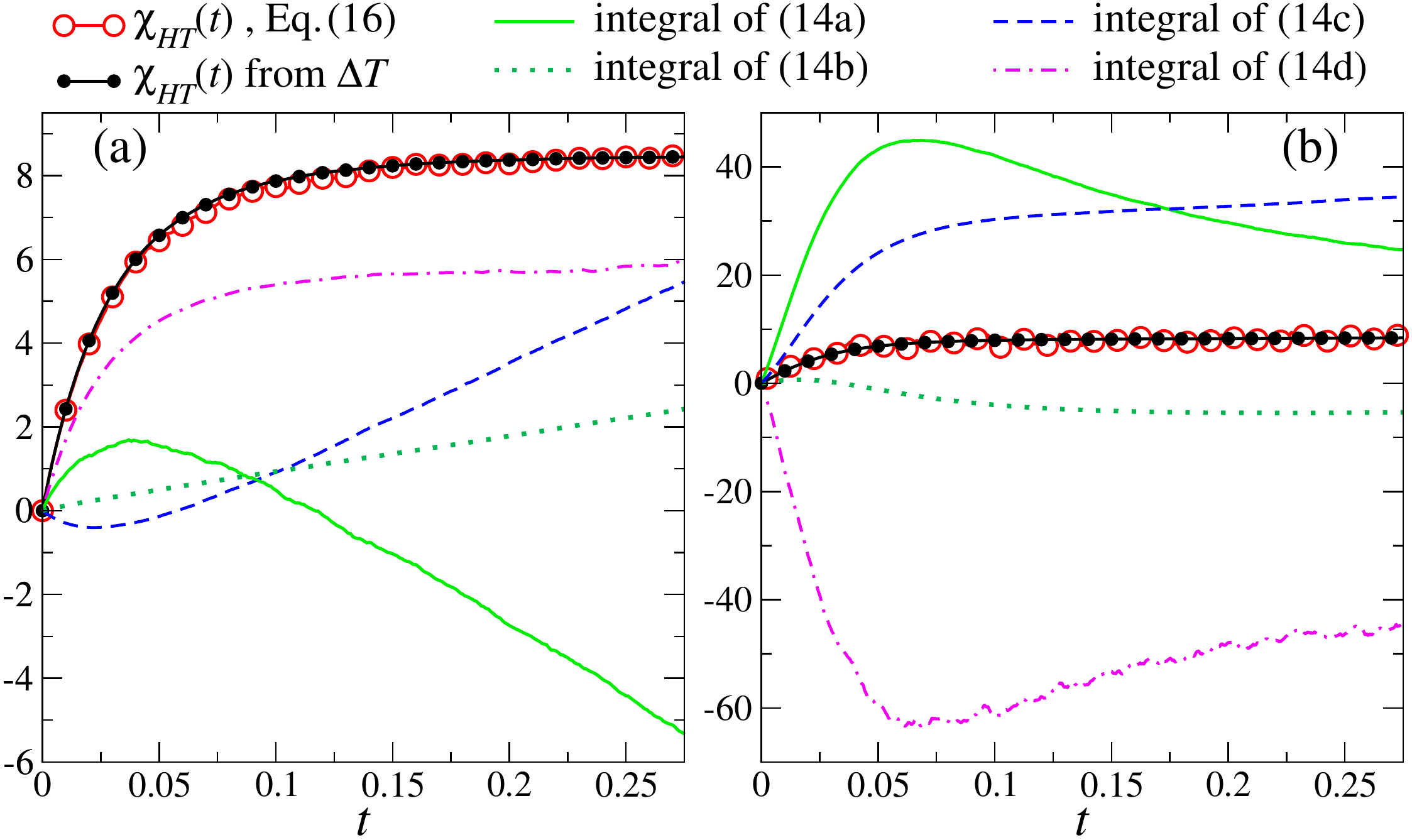}
\caption{(Color online) 
Susceptibility of the internal energy to a change of $T$
in the elastic slab with fixed $T_2\ne T$, 
computed both by the direct perturbation~\eqref{chi2} and through~\eqref{chi}:
(a) steady state at $t=0$, and (b) transient from a configuration given at $t=0$.
Terms of~\eqref{chi} specified in~\eqref{resp} are also shown.
}
\label{fig:chiU} 
\end{figure}

As an example, we show the susceptibility of the energy ($\caO = \caH$) to a change of $T$ in a model of elastic slab between two thermostats, simulated using a Heun scheme~\cite{sek10} so that the points of the trajectory can be used in discretized Stratonovich integrals. A scalar $x_i$ is defined for $i$ labeling a site in a portion $L\times L \times 2$ of a cubic lattice. Each $x_i$ in the lower $L\times L$ layer is thermalized at $T$ while the $x_i$'s in the upper sites are maintained at $T_2\ne T$, so that the system is out of equilibrium due to a constant heat flux.
The total energy is
$\caH = \sum_i  u(x_i) +\sum_{i\div j} u(x_i-x_j)$
with $u(x) = \frac {x^2}2 + \frac {x^4}4$
($i\div j$ indicates the nearest neighbor pairs, with periodic boundary conditions within each layer).
We compute $\chi_{\caH T}(t)$ both by direct application of a constant perturbation $\Delta T = T\cdot 10^{-2}$ turned on at time $t=0$,
\begin{equation}
\label{chi2}
\chi_{\caH T}(t) = \frac{\mean{\caH(t)}_{(T+\Delta T,T_2)} - \mean{\caH(t)}_{(T,T_2)}}{\Delta T},
\end{equation}
and by~\eqref{chi}.
We find that the two estimates of the susceptibility are in good agreement.
For instance, starting from the system in the steady state at $t=0$, 
Fig.~\ref{fig:chiU}(a) shows the results for a slab with $L=5$, when $T=2$, $T_2=3$.
Since here a positive $\Delta T$ brings $T$ closer to $T_2$, in response the mean heat current $\mean{J}$ from the reservoir $\caT$ is lowered, as captured by the asymptotic anticorrelation between $J$ and energy [integral of~\eqref{resp_a} in Fig.~\ref{fig:chiU}(a)]. Hence, unlike in equilibrium, the entropy flux $J/T$ is by itself not sufficient even for determining the global trend of the response. Fig.~\ref{fig:chiU} shows that all terms in \eqref{resp} are relevant.
To remind that the theory is applicable also to transient conditions, in Fig.~\ref{fig:chiU}(b) we show results obtained by starting at $t=0$ from a given configuration with $x_i=1/2$ in the lower layer and $x_i=-1/2$ in the upper one. In a similar way, one might analyze data from a temperature quench as usually done for  models of spins or glasses~\cite{cug94,lip05,lip07,gar09,mer05}.

The response formula \eqref{resp} simplifies when all the reservoirs are at the same temperature previous to the perturbation, because $J^{\rm{int}}=0$ and $\genL^{(T)}= \genL$:
\begin{align}
&R_{\caO T}(t,t') =   \sum_{i\in \caT} \frac{1}{4 T^2}\Big\{ 
2\mean{\caO(t) J_i(t')} +  \frac{1}{2}\partial^2_{t'} \mean{\caO(t)  x_i^2(t')} \nonumber\\
&\qquad+\mean{\caO(t) F^2_i(t')} -\mean{\caO(t) x_i(t') \genL F_i(t')} 
\Big\}.
\end{align}
Moreover, if the system is in a nonequilibrium steady state, $\genL$ can be conveniently expressed in terms of the generator of the dynamics reversed in time,  $\genL^*$, as  $\genL = \genL^* + 2 \vec v \cdot  \nabla_{ \vec x} $, with $\vec v=\vec J/\rho$ the state velocity, 
i.e.~the probability current over the probability density~\cite{che08}. 
Taking for simplicity only one degree of freedom $x$, 
it is possible~\footnote{We exploit the relations
$\frac{1}{2}\partial^2 _{t'}\mean{x^2(t') \caO(t)} = \mean{J(t') \caO(t)} - \mean{[x \dot  F](t') \caO(t)} - 2\mean{[\dot x x](t') [v \partial_x O](t)}$,
$\mean{\caO(t)   [F^2  -  x  \genL F](t') } = \mean{\caO(t)  [J +  x \dot F] (t')}  + 2 \mean{\caO(t) [v(F-x \partial_x F)](t') }$.}
to recast the temperature response in the form
\begin{align}
\label{resp-v}
&R_{\caO T}(t-t')  = -\frac{1}{T^2} \big[ \mean{\caO(t) \dot x(t') F(t')} \\
&\quad + 2\mean{ \partial_x \caO(t)  v(t) \dot x (t') x(t')} + 2 \mean{\caO(t) v(t')f(t')} \big]\nonumber.
\end{align}
In equilibrium $v = 0$ and $F=-\partial_x \caH$, hence only the entropic term $-\frac{1}{T^2} \mean{\caO(t) \dot x(t') F(t')}=\frac{1}{T^2} \frac{d}{d t'}\mean{\caO(t)  \caH(t')} $ survives, and~\eqref{resp-v} correctly reduces to a Kubo formula. The nonequilibrium corrections are the correlations between the observable, the state velocity and the perturbing forces $f$ and $x$. 

In conclusion, for state observables, a fluctuation-response relation to temperature changes has been derived for overdamped systems out of equilibrium, both for transient conditions and for steady states generated by nonconservative forces or by temperature gradients.
The understanding of the response to temperature changes complements the previous approach based on path-weights, where the response to forces was derived~\cite{bai09,bai09b}.
We can thus say that the picture of the linear response for nonequilibrium overdamped systems is almost complete. To fully close the circle, one needs the temperature response for systems with space-dependent noise prefactors, occurring for instance with hydrodynamic interactions. Investigations of these issues should consider a time rescaling, a key procedure in our approach, which leads to the concept of thermal time.

\begin{acknowledgments}
\paragraph{Acknowledgments:}
The authors would like to thank K.~Kroy and C.~Maes for useful discussions and for their comments on the manuscript.
\end{acknowledgments}


%

\end{document}